\newcommand{\qin}{\mbox {$ q_{1}^{\nu}$}}

\newcommand{\al}{\mbox{$\alpha $}}                                             
\newcommand{\aln}{\mbox{$\alpha _{n}$}}                 
                                      
\newcommand{\s}{\mbox{$\sigma $}}                                              
\newcommand{\SI}{\mbox{$\Sigma $}}    
\newcommand{\tS}{\mbox{$\tilde{\Sigma} $}}    
                                             
\newcommand{\xn}{\mbox{$x_{n}$}}
\newcommand{\xns}{\mbox{$x_{n}(\sigma )$}}
\newcommand{\xnsp}{\mbox{$x_{n}(\sigma ')$}}
\newcommand{\xm}{\mbox{$x_{m}$}}
\newcommand{\xms}{\mbox{$x_{m}(\sigma )$}}
\newcommand{\xmsp}{\mbox{$x_{m}(\sigma ')$}}

\newcommand{\ai}{\mbox{$\alpha _{1}$}}

\newcommand{\e}{\mbox{$e^{ik_{0}X}$}}                                          
\newcommand{\qe}{\mbox{$e^{iq_{0}X}$}}                                         
\newcommand{\at}{\mbox{$\alpha _{2}$}}

\documentstyle[12pt]{article}                                                  
\newcommand{\kim}{\mbox {$ k_{1}^{\mu}$}}                                      
\newcommand{\kom}{\mbox {$ k_{0}^{\mu}$}}                                      
\newcommand{\ki}{\mbox {$ k_{1}$}}
\newcommand{\yn}{\mbox {$ Y_{n}$}}                                             
\newcommand{\ym}{\mbox {$ Y_{m}$}} 
\newcommand{\kn}{\mbox {$ k_{n}$}}
\newcommand{\Kn}{\mbox {$ K_{n}$}}
\newcommand{\km}{\mbox {$ k_{m}$}}
\newcommand{\Km}{\mbox {$ K_{m}$}}

\newcommand{\qo}{\mbox {$ q_{0}$}}                                            
\newcommand{\ko}{\mbox {$ k_{0}$}}

\newcommand{\kin}{\mbox {$ k_{1}^{\nu}$}}

\newcommand{\li}{\mbox {$    \lambda_{1}$}}

\newcommand{\lpp}{\mbox {$e^{i\int _{c} \alpha (t)                             
k(t) \partial _{z} X(z+t) dt +ik_{0}X}$}}

\newcommand{\la}{\mbox{$ \lambda $}}                                           
\newcommand{\be}{\begin{equation}}                                             
\newcommand{\br}{\begin{eqnarray}}                                             
\newcommand{\ee}{\end{equation}}                                               
\newcommand{\er}{\end{eqnarray}}                                               
                   
\newcommand{\gvk}{\mbox {$ e^{i\sum _{n }k_{n}Y_{n}}$}}                      
                     
\newcommand{\eln}{\mbox {$ e^{\sum _{n}\lambda _{-n}L_{+n}}$}}

\newcommand{\dsi}{\mbox {$\frac{\partial}{\partial x_{1}}$}}             
\newcommand{\dsic}{\mbox {$\frac{\partial ^{3}}{\partial x_{1}^{3}}$}}

\newcommand{\dsn}{\mbox {$\frac{\partial }{\partial x_{n}}$}}             
           
\newcommand{\dst}{\mbox {$\frac{\partial }{\partial x_{2}}$}}             
        
\newcommand{\dsth}{\mbox {$\frac{\partial }{\partial x_{3}}$}}

\newcommand{\dsit}                                                             
{\mbox {$\frac{\partial ^{2}}{                                           
\partial x_{1}\partial x_{2}}$}}                                               
                                            
\newcommand{\dsnm}                                                             
{\mbox {$\frac{\partial ^{2}}{                                            
\partial x_{n}\partial x_{m}}$}}                                               
\newcommand{\dsii}{\mbox {$\frac{\partial ^{2}}                           
{\partial x_{1}^{2}}$}}                                                        
                                                     
\newcommand{\p}{\mbox {$ \partial$}}

\newcommand{\pp}{\mbox {$ \partial ^{2}$}}                                     
\newcommand{\mup}{\mbox {$ \partial _{\mu}$}}

\begin{document}                                                               
\title{
\hfill\parbox{4cm}{\normalsize IMSC/97/04/13\\
			       hep-th/9704116}\\        
\vspace{2cm}
Loop Variables and Gauge Invariant Interactions in String
Theory\thanks{
Talk given at the Puri Workshop on
Frontiers in Field Theory, Quantum Gravity and String Theory 
in Dec 1996} }                                        
\author{B. Sathiapalan\\ {\em                                                  
Institute of Mathematical Sciences}\\{\em Taramani                     
}\\{\em Chennai, India 600113}}                                     
\maketitle                                                                     
\begin{abstract}                                                               
We describe a method of writing down the exact interacting gauge invariant
equations for all the modes of the bosonic open string. It is
a generalization of the loop variable approach that was used
earlier for the free, and  lowest order interacting cases.
The generalization involves, as before, the introduction of a parameter
to label the different strings involved in an interaction.  The interacting
string has thus becomes a ``band'' of finite width.  As in the free case,
the fields appear to be massless in one higher dimension.  Although
a proof of the consistency and gauge invariance to all orders 
(and thus of equivalence with string theory)
is not yet available, plausibility arguments are given. We also give
some simple illustrations of the procedure.
\end{abstract}                                                                 
\newpage                                                                       
\section{Introduction}                                                         
The loop variable approach introduced in \cite{BS1} (hereafter I)
is an attempt to write
down gauge invariant equations of motion for both massive and massless 
modes.  This method being rooted in the sigma model approach 
\cite{C,S,DS,FT,Polch,T},the  
computations are expected to be simpler and the gauge transformation
laws more transparent.  This hope was borne out at the free level
and also to a certain extent in the interacting case \cite{BS2} (
hereafter II).  The gauge transformations at the free level 
can be summarized
by the equation
\be \label{1.1}
k(t) \rightarrow k(t) \lambda (t)
\ee
Here $k(t)$ is the generalized momentum Fourier-conjugate to $X$ and
$\lambda$ is the gauge parameter.  This clearly has the form
of a rescaling and one can speculate on the space-time interpretation
of the string symmetries as has been done for instance in I. 

In II the interacting case was discussed.  It was shown that the leading
interactions could be obtained by the simple trick of
introducing an additional parameter `$\sigma $' as $k(t) \rightarrow
k(t, \sigma )$, parametrizing different interacting strings.
Thus, for instance, $\kim (\sigma _{1} )\kin (\sigma _{2})$ could stand
for two massless photons when $\sigma _{1} \neq \sigma _{2}$, but when
$\sigma _{1} = \sigma _{2}$ it would represent a massive ``spin 2''
excitation of one string.  The gauge transformations admit a corresponding
generalization
\be           \label{1.2}
k(t, \sigma )\rightarrow k(t,\sigma )\int d\sigma _{1} 
\lambda (t,\sigma _{1})                                  
\ee
It was shown, however that this prescription introduces only the
leading interaction terms.

In this talk we show that there is a natural generalization of this 
construction to include the full set of interactions that one
expects based on the operator product expansion (OPE) of
vertex operators.  We verify the gauge invariance (under 1.2) in
one non trivial equation.  We give arguments why we expect this
to hold for all the equations We do not have a complete proof
of this yet.                          

This paper is organized as follows.  In section II we give a short
review of II and elaborate on the role of the parameter $\s$.  In section
III we describe the generalization referrred to above.  In section IV
we give some examples. Section V contains some concluding remarks.

\newpage
\section{Review}
In I the following expression was the starting point to obtain the
equations of motion at the free level: 
\be     \label{2..1}
A=e^{k_{0} ^{2}\Sigma + k_{n} . k_{0} \dsn \Sigma + \sum _{n,m}k_{n} .k_{m}
(\dsnm  - \frac{\partial}{\partial x_{n+m}})\Sigma +ik_{n} Y_{n}}
\ee

The prescription was to vary w.r.t $\SI$ and evaluate at 
$\Sigma =0$ to get the 
equations of motion.  Here, 
$2 \Sigma \equiv < Y(z)Y(z) >$ and $Y = 
\sum _{n}\aln \frac{\partial^{n}X}{(n-1)!} \equiv
\sum _{n} \alpha _{n} \tilde{Y}_{n} $. 
$\aln $ are the modes
of the einbein $\al (t)$ used in defining the loop variable
\be \label {lv}
\lpp = \gvk
\ee
 One can also show easily that 
$Y_{n} = \frac{\partial Y}{\partial x   _{n}}$.  
$\SI$ is thus a generalization of the Liouville mode, and what we
have is a generalization of the Weyl invariance condition on vertex 
operators.  

There is an alternative way to obtain the $\Sigma $ dependence \cite{BSCD}.
This is to perform a general conformal transformation on a vertex
operator by acting on it with $\eln $ using the relation 
\footnote{This relation is only true to lowest order in
$\lambda$.  The exact expression is given in 
\cite{BSV}}{\cite{BSV}:
\be  \label {2.2.5}
\eln e^{iK_{m} \tilde{Y}_{m}}
= e^{K_{n}.K_{m}\lambda _{-n-m}+\tilde{Y_{n}}\tilde{Y_{m}}\lambda_{+n+m}
+imK_{n}\tilde{Y_{m}}\lambda_{-n+m}}e^{iK_{m}\tilde{Y_{m}}}
\ee
The anomalous term is $\Kn .\Km \lambda _{-n-m}$ and the classical 
term is $ m\Kn \tilde{\ym} \lambda _{-n+m}$. We will ignore the 
classical piece: this can be rewritten as a $(mass)^{2}$ term which 
will be reproduced by performing
a dimensional reduction and other pieces involving derivatives
of $\Sigma$ (defined below) that correspond 
to field redefinitions \cite{BS1}.  
We can apply (\ref{2.2.5}) to the loop variable 
(\ref{lv}) by setting $\Km = \sum _{n} k_{m-n} \aln$.  Defining
\be
\Sigma =  \sum _{p,q} \alpha _{p} \alpha _{q} \lambda _{-p-q}
\ee
we recover (\ref{2.1}). It is the approach described above that 
generalizes more easily to the interacting case.

The equations thus obtained are invariant under 
\be   \label{2.3}
k_{n} \rightarrow \sum _{m}k_{n-m} \lambda _{m}
\ee
which is just the mode expansion of (\ref{1.1}). 

That this is an invariance of the equations of motion 
derived from (\ref{2.1}) follows essentially from the fact 
that the transformation (\ref{2.3}), applied to (\ref{2.1})
changes it by a total derivative.
\be  \label{2.4}
\delta A = \sum _{n} \lambda _{n} \dsn [A]
\ee
 Thus the equations obtained from (\ref{2.1}) cannot be affected.
		      
However there are some caveats.  In proving (\ref{2.4})
one needs to use equations such as
\be \label{2.5}
\dsi (\dsii - \dst )\SI = (\dsic
-\dsit )\SI = 2(\dsit - \dsth ) \SI
\ee
which follow from the basic definitions \cite{BS1}.  This results
in certain constraints having to be satisfied in order
for the equations to be invariant. An example is the tracelessness
constraint 
\be   \label{2.6}
\li k_{1}.k_{1} =0
\ee
described in I.  For further details we refer the reader to \cite{BS1}
and \cite{BS3}.

In II this approach was generalized to include some interactions.  
The basic idea was to introduce a new parameter 
$\sigma : 0\leq \sigma \leq 1$
to label different strings and to replace each $\kn$ in the free 
equation by $\int _{0}^{1} d \sigma \kn (\sigma )$. 
The next step was to assume
that 
\be  \label{2.7}
<\kim (\sigma _{1})\kin (\sigma _{2})> = S^{\mu\nu}\delta 
(\sigma _{1}-\sigma_{2})
+ A^{\mu}A^{\nu}
\ee
where $< ...>$ denotes $\int {\cal D}k(\sigma ) ...\psi [k(\sigma )]$, 
$\psi $ being
the``string field''  defined in I.\footnote{No special property of 
$\psi $ is assumed other than this.} This corresponds to saying that when 
$\sigma _{1} = \sigma _{2}$, both the $\ki $'s belong to the same string
and otherwise to different strings where they represent two 
photons at an interaction point.\footnote{It is conceivable that   
(\ref{2.7}) may have to be generalized by replacing the $\delta$-function
on the RHS by something else, when we go to the fully interacting case.
However in this talk we will not do so - in the example considered
(albeit a very simple one) (\ref{2.7}) seems to be sufficient.}
The gauge transformation is replaced 
by (\ref{1.2}).  This is easily seen to give interacting interacting
equations.  However the fact is that this is only a leading term
in the infinite set of interaction vertices.

As a prelude to generalizing this construction, let us explain
more precisely the nature of the replacement $\kn \rightarrow
\int _{0}^{1} d \sigma \kn ( \sigma )$.  Let us split the interval
$(0,1)$ into $N$ bits of width $a=\frac{1}{N}$.  We will
assume that when $\s$ satisfies $\frac{n}{N} 
\leq \sigma \leq \frac{n+1}{N}$
it represents the $(n+1)$th string.  Let us also define a function
\br \label{2.8} 
D(\sigma _{1},\sigma _{2})& =&1 \;  if\; 
\sigma _{1},\sigma_{2}\; belong\; to\; the\;
same\; interval     \nonumber    \\
&=&0 \;if\;\sigma_{1}\;,\sigma_{2} \;belong\;to\;different\;intervals.
\er
Thus $\int _{0}^{1} d \sigma _{1}D(\sigma _{1} \sigma _{2})  = \; a \; =
\int _{0}^{1} d \sigma _{1}\int _{0}^{1} d 
\sigma _{2}D(\sigma _{1} \sigma _{2})$.

Then we set 
\be   \label{2.9}
<k^{\mu}(\sigma_{1})k^{\nu}(\sigma_{2})> = 
\frac{D(\sigma_{1},\sigma_{2})}{a}S^{\mu \nu}   
+A^{\mu}A^{\nu}
\ee
In the limit $N \rightarrow \infty ,\; a \rightarrow 0, \; 
\frac{D(\sigma _{1} , \sigma _{2})}{a}  
\approx \delta (\sigma _{1} -\sigma _{2})$
and we recover (\ref{2.7}). 

In effect (\ref{2.1}) has been modified to
\be  \label{2.10}
e^{\int_{0}^{1}\int_{0}^{1}  d\sigma_{1}d\sigma_{2}
[k_{0}(\sigma_{1})k_{0}(\sigma_{2})\Sigma
+k_{n}(\sigma_{1}).k_{0}(\sigma_{2})\dsn \Sigma + \sum _{n,m}
(\dsnm - \frac{\partial}{\partial x_{n+m}})\Sigma ] +\int _{0}^{1}
d\sigma k_{n}(\sigma )Y_{n}}
\ee
The final step (which is also necessary in the free case), is to
dimensionally reduce to obtain the massive equations.  For details
we refer the reader to I.

The modification (\ref{2.7}), that replaces $S^{\mu \nu}$ by
$S^{\mu \nu} + A^{\mu} A^{\nu}$ can be understood in terms of
the OPE.  Consider a correlation function involving two vector
vertex operators and any other set of operators, that we represent as
\be  \label{2.11}
{\cal A}=<V_{1}V_{2}...V_{N}:\kim \partial_{z}X^{\mu}\e 
:\qin \partial_{w}X^{\nu}
 \qe>
\ee
The OPE of 
$:\kim \partial_{z}X^{\mu}(z) \e :$ and $: \qin \partial_{w} 
X^{\nu}(w)\qe :$

is given by 
\[
:\kim \partial_{z}X^{\mu}(z) \e :: \qin \partial_{w} X^{\nu}(w)\qe : =
\]
\be  \label{2.12}      
:\kim \qin \partial_{z} X^{\mu} \partial_{w} X^{\nu} 
e^{i(k_{0}X(z) + q_{0}X(w))}:
+\; terms\; involving\; contractions.
\ee

We can Taylor expand
\be     \label{2.13}
X(w) = X(z) + (w-z)\partial _{z} X + O(w-z) +...
\ee

This gives for the leading term in (\ref{2.11})
\be     \label{2.14}
{\cal A} = <V_{1}V_{2}...V_{N}:\kim \qin \partial_{z} X^{\mu} 
\partial_{z} X^{\nu} e^{i(k_{0}X(z) + q_{0}X(w))}:>
\ee
Compare this with the correlation involving $S^{\mu \nu}$:
\be     \label{2.15}
{\cal A'} = <V_{1}V_{2}...V_{N}:\kim \kin \partial_{z} X^{\mu} 
\partial_{z} X^{\nu} \e:>
\ee
We see that ${\cal A} $ and ${\cal A'}$ give identical terms except
that $S^{\mu \nu}$ is replaced by $A^{\mu}A^{\nu}$.  It is 
in this sense that the substitution given in II, gives the leading term 
in the OPE.  The crucial point is that, while in (\ref{2.10}) we have
introduced the parameter $\sigma $ in the $k _{n}$'s we have not 
done so for the $\yn$'s.  This is equivalent to approximating
$X(w)$ by $X(z)$ in (\ref{2.13}).  Clearly, the generalization required
to get all the terms is to introduce the parameter $\s$ in $Y$ also.
We turn to this in the next section.

\section{Introducing $\s$-dependence in the loop variable}

We will introduce the parameter $\s$ in all the variables keeping
in mind the basic motivation that $\s$ labels different vertex 
operators.  Thus all the variables that are required to define
a vertex operator become $\s$ dependent.
Thus
\be     \label{3.1}
X^{\mu}(z)\rightarrow X^{\mu}(z(\sigma ))
\ee
\be     \label{3.2}
\xn \rightarrow \xn (\sigma )
\ee
in addition to
\be     \label{3.3}
\kn ^{\mu}\rightarrow \kn ^{\mu}(\sigma )
\ee
(\ref{3.1}) and (\ref{3.2}) imply that
\be     \label{3.4}
\frac{\partial}   {\partial\xn} Y \rightarrow 
\frac{\partial}{\partial \xns}
Y(z(\sigma ),\xns ) 
\ee
Note that $X$ need not be an explicit function of $\s$ since at a given
location $z$, on the world sheet there can only be one $X(z)$.  As an
example of the above consider the case when we have regions $(0,1/2)$
and $(1/2,1)$.  When $0\leq \sigma \leq 1/2$ one has $z(\sigma ) \equiv z$
and for $1/2 \leq \sigma \leq 1$ one has $z(\sigma ) \equiv w$.  Similarly
$x_{n}(\sigma )$ could be called $x_{n},y_{n}$ in the two regions
and $\kn (\sigma )$ could be called $\kn , p_{n} $ in the two regions.
Thus in this example the vertex operator 
$\kn (\sigma )\yn (z(\sigma ), x_{n}( \sigma ))
e^{ik_{0}(\sigma )Y(\sigma )}$
stands for $\kn \frac{\partial Y}
{\partial x   _{n}}(z,x_{i})e^{ik_{0}Y(z,x_{n})}$
and $p_{n} \frac{\partial Y}{\partial y_{n}}(w,y_{i})
e^{ip_{0}Y(w,y_{n})} $ in the
two regions.

We will further assume that the integration measure of the
free theory $[dx_{1}dx_{2}.....]$ is replaced by
$[{\cal D}x_{1}(\sigma ) {\cal D}x_{2}(\sigma )....]$so that we can 
continue to integrate by parts.  However we have to clarify what we
mean by a derivative w.r.t $x_{n}(\sigma )$:  In (\ref{3.4}) we have 
$\frac{\partial Y(z(\sigma ), x_{i}(\sigma ))}{\partial\xns}$ : 
One has to specify
the meaning of $\frac{\partial\xns }{\partial\xnsp}$.  
Clearly what we want is:
If $\sigma , \sigma '$ belong to the same interval, then 
$\frac{\partial\xns }{\partial\xnsp} \; = \; 1$ and zero otherwise. 
Thus using
(\ref{2.8})
\be     \label{3.5}
\frac{\partial\xns}{\partial\xnsp} = D(\sigma , \sigma ')
\ee
or more generally
\be     \label{3.6}
\frac{\partial\xns}{\partial\xmsp} = \delta _{nm} D(\sigma ,\sigma ')
\ee
Note that this is not the same as the conventional functional
derivative.  However we can define
\be     \label{3.7}
\frac{\delta \xns }{\delta \xmsp } \equiv \frac{D(\sigma ,\sigma ')}{a}
\ee
which, in the limit $a\rightarrow 0$ becomes the usual
functional derivative.  Thus
\be     \label{3.8}
\int d \sigma ' \frac{\delta Y (\sigma )}{\delta \xnsp} 
= \frac{\partial Y(\sigma )}{\p
\xns}
\ee
We can now write down the generalization of (\ref{2.1})
\[
exp \{\int \int d\sigma_{1} d \sigma _{2} 
\{ k_{0}(\sigma _{1}).k_{0}(\sigma _{2})
[\tS (\sigma _{1},\sigma _{2}) + \tilde{G}(\sigma_{1},\sigma _{2})]
\]
\[
+\sum _{n>0}\int d\sigma _{3} \kn (\sigma _{1}).\ko (\sigma _{2}) 
\frac{\delta}   
{\delta\xn (\sigma _{1})}[\tS (\sigma _{3},\sigma _{2}) 
+ \tilde{G}(\sigma _{3},\sigma _{2})]
\]
\[
+\int \int d\sigma _{3} d\sigma _{4}\sum _{n,m >0} 
\kn (\sigma _{1}).\km (\sigma _{2})
\]
\[
\frac{1}{2}[\frac{\delta ^{2}}{\delta \xn (\sigma _{1})
\delta \xm (\sigma _{2})}
-\delta (\sigma _{1}-\sigma _{2})\frac{\delta }
{\delta x_{n+m}(\sigma _{1})}]
[\tS(\sigma _{3},\sigma _{4}) + \tilde{G}(\sigma _{3},\sigma _{4})]\}\}
\]
\be     \label{3.9}  
exp \{i\int d \sigma k_{n}(\sigma )Y_{n}(\sigma )\}
\ee
In (\ref{3.9}) $G(\sigma _{1} ,\sigma _{2})=
\tilde{G}(z(\sigma _{1}), z(\sigma _{2}))$
=$<Y(z(\sigma _{1}))Y(z(\sigma_{2}))>$ is the Green function which starts
out as $ln (z_{1} - z_{2})$.  More precisely, if we define:
[Using the notation $z_{i} = z(\sigma _{i})$]
\be     \label{3.10}
D_{z_{1}} = D_{z(\sigma _{1})} \equiv 1+\ai 
(\sigma _{1})\frac{\partial}
{\partial z (\sigma _{1})} + \at \frac{\pp}
{\partial z^{2} (\sigma _{1})}+...
\ee
so that
\be    \label{3.11}
Y(z(\sigma ))=D_{z(\sigma )}X(z(\sigma ))
\ee
then,
\be     \label{3.12}
\tilde{G}(z_{1},z_{2}) = D_{z_{1}} D_{z_{2}}G(z_{1},z_{2})
\ee
\be     \label{3.13}
\tS(\sigma _{1},\sigma_{2}) =  D_{z_{1}} 
D_{z_{2}}\SI (\sigma _{1},\sigma _{2})
\ee
where 
\be     \label{3.14}
\Sigma (\sigma _{1} , \sigma _{2})=
\frac{\la (z(\sigma _{1}))-\la (z(\sigma _{2}))}
{z(\sigma _{1})-z(\sigma _{2})}
\ee
is the generalization of the usual $\Sigma (\sigma )$
which is equal to $\frac {d \lambda }{d z}$.  The $\tilde {\Sigma}$
dependence in (\ref{3.9}) is obtained by the following step:
\be     \label{3.15}
e^{:\frac{1}{2}\int du \lambda (u) [\partial _{z} X(z+u)]^{2}:}
e^{ik_{n}\frac{\partial}{\partial x_{n}} D_{z_{1}}X}
e^{ip_{m}\frac{\partial}{\partial x_{m}}D_{z_{2}}X}
\ee
defines the action of the Virasoro generators on the two sets of 
vertex operators.
\be     \label{3.16}
= e^{ik_{n}.p_{m}\partial _{x_{n}}\partial _{y_{m}}D_{z_{1}}D_{z_{2}}
\frac{\lambda (u)}{z_{1}-z_{2}}[\frac{1}{z_{1}-u} -\frac{1}{z_{2}-u}]}
\ee
\be     \label{3.18}
=e^{ik_{n}.p_{m}\partial _{x_{n}}\partial _{y_{m}} \tilde{\Sigma}}
\ee     
This expression is only valid to lowest order in $\lambda $
which is all we need here.\footnote{The exact expression is given in
\cite{BSV}}.
The expression 
\be     \label{3.20}
\int \int d\sigma _{1} d\sigma _{2}
\frac{1}{2}[\frac{\delta ^{2}}
{\delta \xn (\sigma _{1})\delta \xm (\sigma _{2})}
-\delta (\sigma _{1}-\sigma _{2})
\frac{\delta }{\delta x_{n+m}(\sigma _{1})}]
[\tS(\sigma _{3},\sigma _{4}) + \tilde{G}(\sigma _{3},\sigma _{4})]
\ee
can easily be seen to be equal to
\be     \label{3.21}
\frac{\pp}{\partial\xn (\sigma _{3}) 
\partial\xm (\sigma _{3})}\tS (\sigma _{3},\sigma _{4})         
\ee
In the limit $\sigma _{3} = \sigma _{4}=
\sigma $ this is just equal to
$1/2[\frac{\pp}{\partial\xns \partial\xms} - 
\frac{\partial}   {\partial x_{m+n}(\sigma )}]
\tilde{\Sigma} (\sigma , \sigma )$ and 
reduces to the free field case described
by (\ref{2.10})(provided the limit is taken after differentiation).

One can show \cite{BSP} that the gauge transformation (\ref{1.2})
changes (\ref{3.9}) by a total derivative
\be     \label{3.22}
\delta A = \int d\sigma \la ( \sigma ) 
\frac{\delta }{\delta \xn (\sigma )}A
\ee
By the argument given in Section II this should mean that the equations
are gauge invariant.  However there is one fact that has to be
kept in mind.  Consider the following expression:
[$\Sigma (\sigma , \sigma )$
is the generalization of the Liouville field that was used in Section II].
\be     \label{3.33}
2(\dsit - \dsth )\SI A + ( \dsii - \dst )\SI 
\frac{\scriptstyle \partial A}
{\scriptstyle \partial x_{1}}
\ee
Using (\ref{2.5}) we get
\be     \label{3.34}
=\; \dsi [(\dsii - \dst )\SI A]
\ee
which is a total derivative.  However if we vary (\ref{3.33})
w.r.t. $\Sigma$, one gets
\be     \label{3.35}
2\delta \SI (\dsit + \dsth )A + 
\delta \SI (\dsii + \dst )  \frac{\scriptstyle \partial A}{
\scriptstyle \partial x_{1}}
\ee
which is not zero. On the other hand if we rewrite (\ref{3.33})
as (using (\ref{2.5}))
\be     \label{3.36}
(\dsic -\dsit )\SI A + (\dsii - \dst )\SI \frac{\scriptstyle \partial A}
{\scriptstyle \partial x_{1}}
\ee
and vary w.r.t $\Sigma$ we get
\be     \label{3.37}
\delta \SI (- \dsic - \dsit ) A + \delta \SI (\dsic + \dsit ) A
\ee
which is zero.

Thus one has to be careful about varying w.r..t $\Sigma$
indiscriminately.  There are certain constraints (such as (\ref{2.5}))
that $\Sigma$ obeys and hence it cannot be varied without taking these 
into account.  In I and II there was a tracelessness condition on the
gauge parameters that saved the day, but this will not always be the
case in the interacting cases considered here.  One possible solution
to the problem (in the example considered above)
is to ensure that the form (\ref{3.36}) is used, rather 
than (\ref{3.33}).  This solution was not available in the free case
because (\ref{3.36}) will generically, lead to higher derivative 
equations of motion, whereas for consistency of field propagation
it is always desirable to have equations that are quadratic in derivatives
- which (\ref{3.33}) implies.  However in interaction terms one can allow
higher derivative terms.  So our solution to this problem will be
to choose the form that ensures gauge invariance.  To be more precise
, constraints such as (\ref{2.5}), imply that $\Sigma$ cannot be 
varied arbitrarily.  Since the constraints depend on $x_{1},x_{2}
x_{3}$ let us asume that $f(x_{1},x_{2},x_{3})$ is a function that
satisfies (\ref{2.5}).  Then $\Sigma (x_{1},...\xn )$ is of the
form $f(x_{1},x_{2},x_{3})\Sigma '(x_{4},..\xn ,...)$.  
Then, if the constraint (\ref{2.5}) is written symbolically as $Df=0$, 
a term of the form
\be     \label{3.38}
Df\delta \SI ' A =0
\ee
can be added to (\ref{3.9}) since it is identically zero.  On the 
other hand, on integration by parts, it adds a piece $D^{\dag}$A
to the equation of motion obtained by varying $\Sigma$.  Thus,
in any equation, for each such constraint, there is an ambiguity 
parametrized by one parameter $x$ of the form $xD^{\dag} A$.  We
have to fix these parameters by requiring gauge invariance.  In the 
examples that we have checked, this is a consistent prescription. We 
do not have an argument that guarantees the consistency in all cases. 

The entity $\tS (\sigma_{1} ,\sigma_{2})$ is a ``bilocal'' field as can
be seen from (\ref{3.13}) and (\ref{3.14}).  In the limit
$\sigma _{1} \rightarrow \sigma _{2}$ 
$\SI (\sigma _{1} ,\sigma_{2}) \rightarrow
\frac{d\lambda}{dz}$, which can be identified with the usual
Liouville mode.  Our strategy will be to Taylor
expand $\SI (\sigma_{1}, \sigma_{2})$ in powers of 
$z(\sigma _{1} ) - z(\sigma _{2}) $,
so that we get $\tS (\sigma _{1} ,\sigma _{1} )$ and derivatives.  In order
to preserve the covariance of the equations of motion we will 
have to use a covariant Taylor expansion of the form (derivation is given 
in \cite{BSP}):
\be     \label{3.39}
\tS (\sigma _{1}, \sigma_{2}) = 
\tS (\sigma _{1}, \sigma_{1}) + [z(\sigma _{2}) -
z(\sigma _{1})] \frac{\partial\tS}
{\partial x_{1}}\mid _{\sigma _{2}=\sigma _{1}}
+O(z(\sigma _{2}) -z(\sigma_{1}))^{2}
\ee
{\em A priori}, $Y$ is a function of $\s$, implicitly through
$z(\sigma)$ and also through $\xn (\sigma)$.  However we will set 
$\xn (\sigma )$
to be independent of $\s$.  This is because gauge transformations
correspond to translations of $\xn (\sigma )$. The gauge parameter
should depend on $z(\s )$ but not directly on $\s$.
Thus we do not lose anything by
setting the $\xn $'s to be equal.  This does not affect earlier 
arguments that show that the gauge variation (\ref{3.9})
is a total derivative in $\xn (\sigma )$ (see (\ref{3.22}).

Finally once the equations obtained from (\ref{3.9}) have been written
down in terms of $\tilde{G} (\sigma _{1},\sigma _{2})$ one has to
actually perform the Koba-Nielsen integration over $z(\sigma _{i})$.
At this point our strategy will be to set all the $\xn $'s to zero,
where upon the integrals reduce to the usual S-matrix type
Koba-Nielsen integration.  The gauge invariance does not
depend on the form of $\tilde{G}$ and hence will continue to
hold.  In actually performing the Koba-Nielsen integrals one will
encounter divergences.  These have to be regulated in the usual way
\cite{BSPT,BSZM}.  Thus we have a prescription
for writing down gauge invariant interacting
equations for all the modes of the string.
In the next section we will illustrate the method
with some simple examples.

\section{Examples}
\setcounter{equation}{0}
\subsection{Tachyon}
We will start by considering the tachyon.  The tachyon, being the
ground state is a little different from the other ones, in that
it is difficult to keep track of how many vertex operator
insertions there are in a term.  Thus the term $e^{i\int_{0}^{1}
d\sigma k_{0} (\sigma ) Y(\sigma )}$ can stand for any number
of exponentials $e^{ik.X}$.  To circumvent this, let us introduce
an identity operator $\int d(\sigma ) {\bf I}(\sigma )$,
which is just equal to $1$ but formally serves the purpose of
counting tachyon insertions.  Thus $\int d \sigma _{1} \int d \sigma _{2}
{\bf I }(\sigma _{1}) {\bf I}(\sigma _{2})$ will signal
the presence of two tachyons. (In all calculations below
 integration over all $\sigma$'s is understood whether or not they
are stated explicitly.  Furthermore integration over $z(\sigma )$ 
(Koba Nielsen integration) is also understood.) And we can let
\be     \label{4.1}
<{\bf I}(\sigma ) \kom (\sigma _{1} )> = 
\mup \phi \frac{D (\sigma ,
\sigma _{1})}{a}
\ee
\be     \label{4.2}
<{\bf I}(\sigma _{1}) {\bf I}(\sigma _{2})> = \phi ^{2}
\ee

In evaluating the action of $\eln $ in (\ref{2.2.5}) we had neglected
the  ``classical'' piece which was really the naive classical dimension - 
the oscillator number.  This was incorporated as a $D+1 ^{th}$
component of momentum \cite{BS1}.  In I it was called $q_{0}$ and we set 
$\qo ^{2} = m^{2}$.  We will continue to do that in the interacting case.
However we have to interpret terms of the form $\qo . p_{0}$ that arise
from $\ko (\sigma _{1} ).\ko (\sigma _{2})$ when 
$\sigma _{1} \neq \sigma _{2} $.
It turns out that this can be given a natural interpretation as 
 the sum of the oscillator numbers of the oscillators that
have been contracted.  It thus gives $(z _{1} -z_{2})^{k_{0}(\sigma _{1})
.k_{0}(\sigma _{2})+q_{0}.p_{0}}$.  Similarly in the coefficient of $\Sigma
(\sigma _{1} , \sigma _{2})$ we have a term $\qo .p_{o}$.  Here again it
will stand for the number of contractions.  But we add one because 
$dz_{1}$, which has a dimension one, also contributes to the Liouville
mode dependence.
  Thus $\Sigma (\sigma _{1}, \sigma _{2})$ measures the dimension
not only of the derivatives ``$\partial_{z}$" in the correlation 
but also the dimension of $dz_{1}$.  Having incorporated this prescription
we can treat all particles as massless, just as in the free case.
Let us proceed with the tachyon.
\[
\int d\sigma _{1} {\bf I}(\sigma_{1})
e^{i\int d\sigma _{3} d\sigma _{4} k_{0}(\sigma _{3}).k_{0}(\sigma _{4})
[\tilde{\Sigma} (\sigma _{3},\sigma _{4}) + 
\tilde{G}(\sigma_{3},\sigma_{4})]}
e^{i\int d\sigma k_{0}(\sigma ) Y(\sigma )} +
\]
\be     \label{4.3}
\int \int d \sigma _{1} d\sigma _{2} \frac{1}{2!}
{\bf I}(\sigma_{1}){\bf I}(\sigma_{2})
e^{i\int d\sigma _{3} d\sigma _{4} k_{0}(\sigma _{3}).k_{0}(\sigma _{4})
[\tilde{\Sigma} (\sigma _{3},\sigma _{4}) + 
\tilde{G}(\sigma_{3},\sigma_{4})]}
e^{i\int d\sigma k_{0}(\sigma ) Y(\sigma )}
\ee
are the first two terms.  The first term gives, using
\be     \label{4.4}
<\ko (\sigma _{4}).\ko (\sigma_{3}){\bf I}(\sigma_{1}) >=
\frac{D(\sigma_{4},\sigma_{1})}{a}
\frac{D(\sigma_{3},\sigma_{1})}{a}\ko ^{2} \phi (\ko )
\ee
and $\tilde{G}(\sigma ,\sigma )=0$ due to normal ordering.
\be     \label{4.5}
\ko (\sigma _{1}).\ko (\sigma_{1}) \phi (\ko ) = 
(\ko ^{2} + \qo ^{2}) \phi (\ko )
= (\ko ^{2} -2 ) \phi (\ko )
\ee
where $\qo ^{2}$ has been set to $-2$.
The interacting piece gives:
\[
\frac{1}{2!}\phi \phi \int \int [
\frac{D(\sigma _{3},\sigma _{1})}{a}  
\frac{D(\sigma _{4},\sigma _{1})}{a} +
\frac{D(\sigma _{3},\sigma _{2})}{a}  
\frac{D(\sigma _{4},\sigma _{2})}{a} +
\]
\[
2\frac{D(\sigma _{3},\sigma _{1})}{a}  
\frac{D(\sigma _{4},\sigma _{2})}{a} ]
1/2\ko (\sigma _{3}).\ko (\sigma _{4})
\tS (\sigma _{3},\sigma _{4})
\]
\be     \label{4.6} 
e^{1/2\int \int d\sigma_{5} d\sigma _{6} 
k_{0}(\sigma _{5})k_{0}(\sigma _{6})
\tilde{G}(\sigma _{5},\sigma _{6})} e^{i\int kY}
\ee
\be     \label{4.7}
=1/4 \phi (\sigma _{1})\phi (\sigma_{2})(\ko + p_{0})^{2}
\tS (\sigma_{1},\sigma_{1})
\int e^{k.pG(\sigma _{1},\sigma_{2})}dz(\sigma_{2})e^{i(k+p)Y}
\ee
We have used (\ref{4.1}) in the exponent also.  Further we have
Taylor expanded $Y(\sigma _{2})$, $\Sigma (\sigma _{2},\sigma_{2})$ and
$\Sigma (\sigma _{1},\sigma_{2})$ around $\sigma _{1}$, and have kept
the lowest term.

Varying wrt $\Sigma $ gives
\be     \label{4.8}
1/4(k+p)^{2}\phi (k)\phi (p) \int dz_{2} (z_{1}-z_{2})^{k.p}
e^{i(k+p)Y}
\ee
Set $q.p =1$ (for $\int dz_{2}$) and $q^{2} = p^{2}=-2$ in
$(k+p)^{2}$ which multiplies $\Sigma $.  Also in
$(z_{1}-z_{2})^{k.p +q.p }$ we set $q.p=0$ (since there are
no contractions) as per the above prescription.  The net result is
\be     \label{4.9}
1/4\frac{(k+p)^{2}}{k.p+1} \phi (k)\phi (p)e^{i(k+p)Y}
\ee
In the on-shell limit, $(k+p)^{2}=2k.p+2$ and we get the quadratic term
in the Tachyon equation \cite{DS}.  Of course, in this problem, 
since no gauge invariance is involved, all this seems 
longwinded.  However it will be necessary with the other modes where
gauge invariance
is involved.

\subsection{Vector}

For the vector, we keep terms of total dimension $\leq 2$. This gives:
\[
exp \{
\frac{\ko (\sigma _{1}).\ko (\sigma _{2})}{2}
[\tS (\sigma _{1},\sigma _{2}) + \tilde{G}(\sigma _{1}, \sigma _{2})] +
k_{1}(\sigma_{1}).k_{0}(\sigma _{2})
\frac{\partial}   {\partial x_{1}(\sigma _{1})}
[\tS (\sigma _{1},\sigma _{2}) + 
\tilde{G}(\sigma _{1}, \sigma _{2})] +
\]
\[
\frac{k_{1}(\sigma _{1}).k_{1}(\sigma _{2})}{2}
\frac{\pp}{\partial x_{1}(\sigma_{1})\partial x_{1}(\sigma_{2})}
[\tS (\sigma _{1},\sigma _{2}) + 
\tilde{G}(\sigma _{1}, \sigma _{2})] +
\]
\be     \label{4.10} 
k_{2}(\sigma _{1}).\ko (\sigma_{2}) \frac{\partial}   
{\partial x_{2}(\sigma_{1})}
[\tS (\sigma _{1},\sigma _{2}) + \tilde{G}(\sigma _{1}, \sigma _{2})]\}
e^{i\int k_{0}(\sigma ) Y(\sigma )}
\ee
We emphasize that the dimensional reduction and identification 
of $\qo ^{2}$ with the massses of the particles is necessary for 
making contact with string theory. None of that is necessary for 
verifying the gauge invariance of the equations obtained from
(\ref{4.10}).  This is in fact a `mysterious' feature of this
construction.  

In this talk we will only consider the simplest interacting example
with two vectors contributing to the vector equation
of motion.  Now, for Abelian fields there are no quadratic 
corrections to the 
equations at this order in derivatives. Thus the answer
we are looking for is in fact zero.  In this sense the calculation
is trivial.  However the vanishing of the quadratic corrections
happens only after Bose symmetrizing on the vector fields,
whereas, gauge invariance
of the equation can be verified even before this is done.  In this sense,
we have a non trivial test of gauge invariance.

At the free level we get:
\be     \label{4.11}
\int \int \frac{1}{2}\ko (\sigma _{1}).\ko (\sigma_{2})
\tS (\sigma_{1},\sigma_{2})
i\int \ki (\sigma ) Y_{1} + \ki (\sigma _{1}).\ko (\sigma_{2}) 
\frac{\partial}   {\partial x_{1}(\sigma_{1})}
[\tS (\sigma_{1},\sigma_{2})]    e^{i\int k_{0}Y}
\ee
Varying w.r.t $\SI$, after Taylor expanding using,
\[
\frac{\partial}   {\partial x_{1}(\sigma_{1})}
[\tS (\sigma_{1},\sigma_{2})]=
\frac{1}{2}
\frac{\partial}   {\partial x   _{1}(\sigma_{1})}
[\tS (\sigma_{1},\sigma_{1})] + 
O(z(\sigma_{1})-z(\sigma_{2})) +...   
\]
we get for (\ref{4.11})
\be     \label{4.12}
\int \int d\sigma _{1}d\sigma _{2} 
\frac{1}{2}\ko (\sigma _{1}).\ko (\sigma_{2}) i\int \ki (\sigma ) Y_{1}
-
\frac{1}{2}\ki (\sigma _{1}).\ko (\sigma_{2})\int d\sigma i\ko (\sigma ) 
\frac{\partial Y(\sigma )}{\partial x   _{1}(\sigma _{1})} 
\ee
Since $x_{1}(\sigma _{1})$ is independent of $\sigma_{1}$,  
$\frac{\partial Y(\sigma )}{\partial x   _{1}(\sigma _{1})} \equiv 
\frac{\partial Y(\sigma )}{\partial x   _{1}}= Y_{1}(\sigma )$.
Thus we get Maxwell's equations:
\be     \label{4.13}
i(\ko ^{2} \kim - \ki .\ko \kom ) =0
\ee
We now turn to the interacting case.
Terms involving two $k_{1}$'s or one $k_{2}$ are: 
\[
i)\; \; 
1/2[\int \int 
\ki (\sigma_{1}).\ko (\sigma _{2}) \frac{\partial}   
{\partial x   _{1}(\sigma_{1})}
(\tS + \tilde{G})]^{2} 
e^{1/2\int \int k_{0}(\sigma _{3}).k_{0}(\sigma _{4})
(\tilde{G} + \tilde{\Sigma})}e^{i\int k_{0}Y} 
\]
\[
ii)\; \;
1/2\int \int \ki (\sigma_{1}).\ki (\sigma _{2}) 
\frac{\pp}{\partial x   _{1}(\sigma_{1})
\partial x   _{1}(\sigma_{2})}
(\tS + \tilde{G}) 
e^{1/2\int \int k_{0}(\sigma _{3}).k_{0}(\sigma _{4})
(\tilde{G} + \tilde{\Sigma})}e^{i\int k_{0}Y} 
\]
\[
iii) \; \; 
\int \int k _{2} (\sigma_{1}).\ko (\sigma _{2}) 
\frac{\partial}   {\partial  x_{2}(\sigma_{1})}
(\tS + \tilde{G}) 
e^{1/2\int \int k_{0}(\sigma _{3}).k_{0}(\sigma _{4})
(\tilde{G} + \tilde{\Sigma})}e^{i\int k_{0}Y} 
\]
\[
iv)\; \;
\int \int \ki (\sigma_{1}).\ko (\sigma _{2}) \frac{\partial}   
{\partial  x_{1}(\sigma_{1})}
(\tS + \tilde{G}) 
e^{1/2\int \int k_{0}(\sigma _{3}).k_{0}(\sigma _{4})
(\tilde{G} + \tilde{\Sigma})}e^{i\int k_{0}Y} \int i \ki Y_{1}
\]
\[
v)\; \;
e^{1/2\int \int k_{0}(\sigma _{3}).k_{0}(\sigma _{4})
(\tilde{G} + \tilde{\Sigma})}e^{i\int k_{0}Y} \int i k _{2} Y_{2}
\]

At the interacting level $k_{2}^{\mu}$ also transforms into
$A_{\mu}$ \cite{BS2}, therefore we need to include terms involving
$k_{2}^{\mu}$ for gauge invariance.  Consider the terms contributing
to $Y_{1}^{\mu}$.  Taylor expanding $\tS$ we find:
\[
i)\; \; [(\frac{1}{2}\ki (\sigma_{1}).\ko (\sigma _{2}) 
+ 
\frac{1}{2}\ki (\sigma_{2}).\ko (\sigma _{1})) 
\frac{1}{2}\frac{\partial}   {\partial  x_{1}(\sigma_{1})} 
\tS (\sigma_{1},\sigma _{1}) ]
\]
\be     \label{4.15} 
[\ki (\sigma_{3}).\ko (\sigma_{4})
\frac{\partial  \tilde{G} (\sigma_{3},\sigma_{4})}
{\partial  x_{1}(\sigma_{3})}]e^{i\int k_{0}Y}
\ee
Varying w.r.t $\SI$
\be     \label{4.16}
i)\; \; -1/4[\ki (\sigma_{1}).\ko (\sigma _{2}) + 
\ki (\sigma_{2}).\ko (\sigma _{1})]
\ki (\sigma_{3}).\ko (\sigma_{4})
\frac{\partial  \tilde{G} (\sigma_{3},\sigma_{4})}
{\partial  x_{1}(\sigma_{3})}e^{i\int k_{0}Y}
i\int \ko Y_{1}
\ee
Using
\[
\frac{\scriptstyle \partial ^{2} \SI (\sigma _{1},\sigma _{2})}
{\scriptstyle \partial x_{1}(\sigma_{1})\partial x_{1}(\sigma_{2})}
=\frac{1}{2}(\dsii -\dst )\tS (\sigma_{1},\sigma_{2}) +
O(z(\sigma_{2})-z(\sigma_{2}))
\]
\be     \label{4.17}
ii)\; \; 1/2 \ki (\sigma_{1}).\ki (\sigma _{2}) \frac{\pp}
{\partial  x_{1}(\sigma_{1})
\partial  x_{1}(\sigma_{2})}
(\tS + \tilde{G}) 
e^{1/2\int \int k_{0}(\sigma _{3}).k_{0}(\sigma _{4})
(\tilde{G} + \tilde{\Sigma})}e^{i\int k_{0}Y} 
\ee
Varying w.r.t. $\SI$ gives:
\be     \label{4.18}
ii)\;\; 1/4\ki (\sigma_{1}).\ki (\sigma _{2}) k_{0}(\sigma _{3}).k_{0}(\sigma _{4})  
\frac{\scriptstyle \partial \tilde{G}}{\scriptstyle \partial
x_{1}}i\int \ko Y_{1}
\ee
(iii) and (v) do not contribute to $Y_{1}$.  (iv) gives
\[
iv)\; \; \ki (\sigma_{1}).\ko (\sigma _{2})
\frac{\partial}   {\partial  x_{1}(\sigma_{1})} \tilde{G} 
\frac{1}{2}
k_{0}(\sigma _{3}).k_{0}(\sigma _{4}) \tS(\sigma_{3},\sigma_{4}) 
\int i \ki Y_{1}
\] 
\[
 + 1/2[\ki (\sigma_{1}).\ko (\sigma _{2}) +  
 \ki (\sigma_{2}).\ko (\sigma _{1})]  
1/2\frac{\partial}   {\partial  x_{1}(\sigma _{1})}\tS  
\]
\be     \label{4.19}
e^{1/2\int \int k_{0}(\sigma _{3}).k_{0}(\sigma _{4})
\tilde{G}}e^{i\int k_{0}Y} \int i \ki Y_{1}
\ee
Varying w.r.t. $\SI$ gives
\[
iv)\;\; 1/2k_{0}(\sigma _{3}).k_{0}(\sigma _{4}) 
\ki (\sigma_{1}).\ko (\sigma _{2})
\frac{\partial \tilde{G}(\sigma _{1}, \sigma _{2})}{\partial x_{1}}
\int i \ki Y_{1}
\]
\be     \label{4.20}
- 1/4[\ki (\sigma_{1}).\ko (\sigma _{2}) +  
\ki (\sigma_{2}).\ko (\sigma _{1})]  
1/2 k_{0}(\sigma _{3}).k_{0}(\sigma _{4}) 
\frac{\partial \tilde{G}}{\partial x_{1}} 
i\int \ki Y_{1}
\ee
The final result, (\ref{4.16}) + (\ref{4.18}) + (\ref{4.20}) 
is manifestly gauge
invariant under
\[
k_{1} (\sigma _{1}) \rightarrow \ko (\sigma _{1}) \int d\sigma           
\li (\sigma )
\] 
Note that the expressions are multiplied by
$e^{k_{0}(\sigma _{i} ).k_{0}(\sigma _{j}) 
\tilde {G}(\sigma _{i},\sigma _{j})}$ which is just $(z(\sigma _{i})
-z(\sigma _{j}))^{k_{0}(\sigma _{i}).k_{0}(\sigma _{j})}$ once we 
set $\xn = 0$. 
   
To get the final answer in terms of space-time fields, and to get their
gauge transformation laws one has to use (\ref{2.9}).  We will
not work out the details here.  In the present 
example because of the antisymmetry of $z(\sigma _{1})-z(\sigma _{2})$
under $\s_{1} \leftrightarrow \s_{2}$,
it 
is clear that the final answer cannot be symmetric under exchange
of the two photons and therefore the final answer is zero.  
This is just a reflection of the fact that $F^{\mu \nu}F^{\nu \rho}
F^{\rho \mu}$ is identically zero for Abelian fields and there is no  
quadratic correction to Maxwell's equations.  
Of
course if we consider higher derivatives or if there are group 
indices on the gauge fields this will not
be true.

There is one remark that needs to be made.  At first sight it is
a little puzzling that the gauge transformation law is so simple
in an interacting theory, and furthermore that it is the
same in the fully interacting version described in this talk as it was
in the earlier case (see section II of this talk) \cite{BS2}.  The resolution of this puzzle
is that the gauge transformation law expressed in terms
of {\em spacetime fields} is more complicated 
in the interacting case. For e.g. the transformation of 
$\mup S^{\mu \nu}$ as given
by the rules described above,
cannot be obtained by differentiating that of $S^{\mu \nu}$ given in II.!  
Thus, the exact transformation law of $S^{\mu \nu}$
is written as an expansion in powers of momentum and has to be
worked out by considering 
the transformations of $\ki \ki , \ki \ki \ko , \ki \ki \ko \ko ....$
In other words, the 
gauge transformation law of $S^{\mu \nu}$ , or any other field, as
given by the law (\ref{1.1}) is only exact for constant fields, or
equivalently at any one space-time point. One has to apply the
prescription (\ref{1.1}) and (\ref{2.7}) to the derivatives
separately.  One can combine these to get a Taylor expansion
that can be worked out to the desired degree of accuracy.
These details and other examples of interacting equations will
be presented elsewhere.

\section{Conclusions}

We have described a general construction that gives gauge
invariant equations of motion, the gauge transformation prescription
(in terms of loop variables) being the same as in II. The main 
advantages are that the prescription for writing
down the equations and gauge transformation laws are fairly
straightforward.  The gauge transformations written in terms of
loop variables seem to have some geometric meaning - they look
like local scale transformations.  The interactions look
as if they have the effect of converting a string to a membrane.
The fields also appear massless in one higher dimension.
These are intriguing features.
There is
no requirement of being on-shell, in this method. It should therefore  
be possible to show that it is equivalent to (open) string field theory.
\cite{SZ,BP,W}

There are several questions that
need to be answered before one can claim that these are the 
interacting (tree level) string equations.  One is the resolution
of the ambiguity described in Section III.  It has to be checked whether 
the dimensional reduction works in all cases.  Formally since the
calculations are identical on-shell with the S-matrix calculation,
one should not have any problems. Whether problems arise off-shell,
needs to be checked. 
As shown in \cite{BSFC}, even U(1) gauge invariance of the massless
vector is violated when a finite cutoff is introduced in order to go  
off-shell, and one needs to introduce massive modes to restore gauge  
invariance.  In the loop variable formalism all the modes
are present from the start and there should not be a problem.
Gauge invariance seems to be exact, on or off-shell.
However the exact value of the Koba-Nielsen integral will depend
on the cutoff prescription.  This issue needs to be clarified.  Also,
one needs to prove that in all cases, and to all orders, the 
transition to space-time fields from the loop variable representation
can be made unambiguously.  Finally, assuming the above issues are 
resolved satisfactorily, one has to see whether this formalism 
provides any insight into the various other issues that have become
pressing in string theory, such as duality.

{\bf Acknowledgements}

I would like to thank the organizers for a very
stimulating and enjoyable workshop.

\end{document}